\documentclass[useAMS,usenatbib]{mn2e}
\usepackage{rotating}
\usepackage{lscape}
\title[Chemical composition of clump stars in NGC~6134]
      {Chemical composition of clump stars in the open cluster NGC~6134
      \thanks{Based on observations collected at ESO telescopes under 
programmes 65.N-0286 and in part 169.D-0473}}

\author[\v{S}ar\={u}nas Mikolaitis et al.]
       {\v{S}ar\={u}nas Mikolaitis,$^{1}$
       Gra\v{z}ina Tautvai\v sien\. e,$^{1}$ 
       Raffaele Gratton,$^{2}$
       Angela Bragaglia,$^{3}$ 
       \newauthor Eugenio Carretta$^{3}$\\ 
$^{1}$Institute of Theoretical Physics and Astronomy, Vilnius University, Go\v{s}tauto 
12, Vilnius 01108, Lithuania \\(e-mail: sarunas.mikolaitis@tfai.vu.lt)\\
       $^{2}$INAF - Osservatorio Astronomico di Padova, Vicolo dell'Osservatorio 5, I-35122 Padova, Italy\\
       $^{3}$INAF - Osservatorio Astronomico di Bologna, Via Ranzani 1, I-40127 Bologna, Italy}
       
\begin{document}

\date{Accepted 2010 .....; Received 2010 .....; in original form 2010 ......}

\pagerange{\pageref{firstpage}--\pageref{lastpage}} \pubyear{2010}

\maketitle

\label{firstpage}

\begin{abstract}
We present an analysis of high-resolution spectra of six core-helium-burning 
`clump' stars in the open cluster NGC~6134. 
Atmospheric parameters ($T_{\rm eff}$,  log~$g$,  $ v_{\rm t}$, and [Fe/H]) were
determined in our previous  study by Carretta et al.\ (2004). In this study we
present abundances of  C, N, O and up to 24 other chemical elements.  Abundances
of carbon were  derived using the ${\rm C}_2$ Swan (0,1) band head at
5635.5~{\AA} (FEROS spectra) and the ${\rm C}_2$  Swan (1,0) band head at
4737~{\AA} (UVES spectra). The wavelength interval 7980--8130~{\AA}, with strong
CN features, was analysed in order to determine  nitrogen abundances and
$^{12}{\rm C}/^{13}{\rm C}$  isotope ratios.  The oxygen abundances were
determined from the [O\,{\sc i}] line at 6300~{\AA}.  Compared with the Sun and
other dwarf stars of the Galactic disk, mean abundances in the investigated
clump stars  suggest that carbon is depleted by about 0.2~dex, nitrogen is
overabundant by about  0.3~dex and oxygen is underabundant by about 0.1~dex. 
This has the effect of lowering the mean C/N ratio to $1.2\pm0.2$.  The mean
$^{12}{\rm C}/^{13}{\rm C}$ ratios are lowered  to  $9\pm2.5$.   Concerning
other chemical elements, the analysis of sodium and magnesium lines (in NLTE), lines of 
other $\alpha$-elements, iron-group and heavier chemical elements gave abundance ratios 
close to the solar ones.
\end{abstract}


\section{Introduction}

Open clusters are important tools for the study of the Galactic disk as well as 
for understanding stellar evolution (e.g.,  Friel et al. 2002; Bragaglia et al.
2008;  Jacobson et al. 2009; Santos et al. 2009). They are the best tool to
understand whether and  how the slope of the radial metallicity distribution
changes with time, since they have formed at all epochs  and their ages,
distances and metallicities are more accurately derived than for the  field
stars. Open clusters are excellent laboratories for investigations of  stellar
evolution as well. Since cluster members were initially of approximately identical chemical
composition, all changes in stellar atmospheres of evolved stars  are related to
internal and external processes of stellar evolution (see, e.g., Pallavicini 2003 and 
references therein). Changes of the abundances of carbon, nitrogen and oxygen are most 
often seen in  evolved stars.  The enhancement of CN bands and altered carbon isotope 
ratios in evolved stars of open clusters were already reported 30 years ago (e.g. Pagel
1974;  McClure 1974). However, the detailed analyses of abundances in stars of
open  clusters from high-resolution spectra are still necessary for
understanding the  processes of dredge-up and extra-mixing affecting the
chemical composition of  atmospheres in evolved low-mass stars. Detailed
spectral analyses of CNO elements in  stars of open clusters are still rather
scarce (Gilroy 1989; Gilroy \& Brown 1991;  Luck 1994; Gonzalez \& Wallerstein
2000; Tautvai\v{s}ien\.{e} et al.\ 2000, 2005;  Origlia et al. 2006; Gratton et
al.\ 2006; Smiljanic et al.\ 2009, etc.). 

The Bologna Open Cluster Chemical Evolution project (BOCCE) is dedicated to
constraining the formation and chemical evolution of the Galactic disk 
(Bragaglia \& Tosi 2006, Carretta et al.\ 2007, and references therein).  This
paper is a part of the effort to determine detailed elemental abundances in open
clusters, with the final goal of deriving the time evolution of abundances in
the Galactic disk. In particular, we concentrate here on carbon and nitrogen,
but also derive abundances of more than 20 other chemical elements.
  
The open cluster NGC~6134 is an intermediate-age, moderately concentrated open cluster 
(Trumpler class II3m) located almost on the galactic plane  
($\alpha_{2000}=16^{h}27.8^{m}, \delta_{2000}=-49^{\circ}09.4^{\prime}.9; l = 324.91^{\circ}, 
b = -0.20^{\circ}$). 
The first extensive photometric study was published by Lindoff (1972) who derived a colour excess 
$E(B-V)=0.45$, a distance of about 700~pc, and an age of about 0.7~Gyr. These 
values were based only on $UBV$ photographic data. Kjeldsen \& Frandsen (1991) 
published $UBV$ CCD data for 66 stars at the center of the cluster and obtained 
$E(B-V)=0.46$ and $m-M=11.25$. 
Coravel radial velocity measurements and photometry in the $UBV$ and $CMT_{1}T_{2}$ system 
of 24 red giants, supplemented by $DDO$ observations of 11 stars, were carried out by 
Claria \& Mermilliod (1992) for membership and binarity analysis, who identified 17 red 
giant members and 6 spectroscopic binaries. The mean cluster radial velocity was found to be 
 $-26.0\pm 0.24$~km~s$^{-1}$, the reddening  $E(B-V)=0.35\pm 0.02$, and the distance 
about 760~pc. The weighted mean value of ${\rm [Fe/H]}=-0.05\pm 0.12$ was evaluated from the 
UV excesses. 
Str\"{o}mgren photometry was analysed by Bruntt et al.\ (1999). 
They determined $E(b-y)=0.263\pm 0.004$ ($E(B-V)=0.365$), [Fe/H]$=0.28\pm 0.02$, and  
age=$0.69\pm 0.10$~Gyr.  
The colour-magnitude diagram (Bruntt et al.\ 1999) shows a "clump" of core-He-burning stars,  
several red giant branch (RGB) stars, and a main sequence. From $BVRI$ CCD observations Ahumada (2002) 
has determined a colour excess 
$0.29 < E(B-V) < 0.37$, age of 1.26~Gyr, and a distance of about $1080\pm 50$~pc, 
which is larger than in the previous analyses.        

\input epsf
\begin{figure}
\epsfxsize=\hsize 
\epsfbox[0 0 500 430]{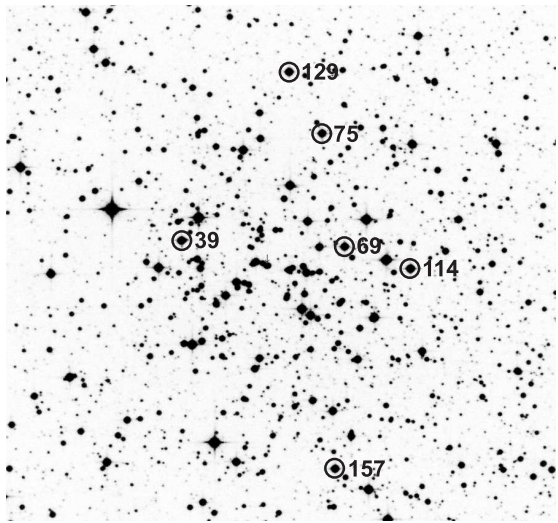}  
\caption{Field of $10\times10$~arcmin$^2$ centered on NGC 6134, with the programme stars indicated 
by their numbers according to Lindoff (1972).}
\label{fig1}
\end{figure}

  
\input epsf
\begin{figure}
\epsfxsize=\hsize 
\epsfbox[-5 -10 600 550]{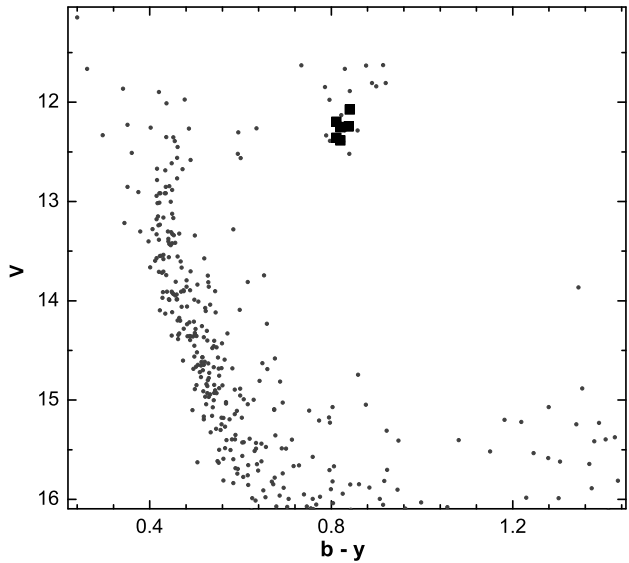} 
\caption{The colour-magnitude diagram of the open cluster NGC~6134. The core-He-burning 
`clump' stars analysed are indicated by the 
    filled squares. The diagram is based on Str\"{o}mgren photometry by Bruntt et al.\ (1999)} 
\label{fig1}
\end{figure}
    
  
Precise iron abundances from high resolution spectra for six stars in this 
cluster have been determined by Carretta et al.\ (2004). An overall metallicity 
${\rm [Fe/H]}=0.15\pm 0.03$ with $rms=0.07$ was found.  Carretta et al. also computed the reddening 
from the temperatures -derived spectroscopically- and the Alonso et al. (1999) colour-temperature 
relations, finding $E(B-V)=0.363\pm0.014$, in very good agreement with Bruntt et al.\ (1999).
Using the same spectra and method of 
analysis, in this paper we continue detailed abundance investigations for the clump stars 
NGC 6134\_39, 69, 75, 114, 129, and 157. 
Elemental abundances in three other clump stars of this cluster 
(NGC 6134\_30, 99, and 202) have been investigated recently by Smiljanic et al.\ (2009). 

\section{Observations and method of analysis}

The spectra of three cluster stars (NGC 6134\_39, 114, and 157) were obtained with 
the spectrograph FEROS (Fiber-fed Extended Range Optical 
Spectrograph) mounted at the 1.5~m telescope in La Silla (Chile). The resolving power is 
$R=48\,000$ and the wavelength range is $\lambda\lambda$ 3700--8600\,{\AA}.  
The stars NGC 6134\_69, 75, and 129 were observed with the UVES spectrograph 
(UV-Visual Echelle Spectrograph) on the Unit~2 of the VLT ESO-Paranal telescope. 
The spectral coverage is  $\lambda\lambda$ 3560--4840\,{\AA},  5550--9460\,{\AA}
and  the resolving power is $R=43\,000$. 
More details can be found in Carretta et al.\ (2004).

Fig.~1 shows a map of the observed stars 
and their evolutionary status  is indicated by their position in 
Fig.~2. All the stars belong to the red clump of the cluster. The log of observations and 
S/N are presented in the paper by Carretta at al.\ (2004). In the same paper all the 
main atmospheric parameters for the observed stars were determined. For convenience 
we present them in this paper as well (Table~1). The effective temperatures were derived 
by minimizing the slope of the abundances from neutral Fe\,{\sc i} lines with respect to the 
excitation potential. The gravities (log~$g$) were derived from the iron ionization 
equilibrium. The microturbulent velocities were determined assuming a relation between 
log~$g$ and $v_t$. The ATLAS models with overshooting were used for the analysis. 
Fe\,{\sc i}  lines were restricted to the spectral range 5500--7000~\AA\ in order to minimize 
problems of line crowding and difficulties in the continuum tracing blueward.  For more details 
and error estimates, see Carretta et al.

   \begin{table}
      \caption{Atmospheric parameters of the programme stars.}
    \[
         \begin{tabular}{rccccrc}
            \hline
	    \hline
            \noalign{\smallskip}
 Star$^*$ & $V$ & $b-y$ & $T_{\rm eff}$ & log~$g$ & [Fe/H] &  $ v_{\rm t}$ \\ 
          &    &   &       K            &         &        &  ${\rm km~s}^{-1}$ \\
            \noalign{\smallskip}
            \hline
            \noalign{\smallskip}
  39 & 12.20 & 0.811  & 4980 & 2.52 & +0.24 & 1.17 \\   
  69 & 12.36 & 0.811  & 4950 & 2.83 & +0.11 & 1.13 \\
  75 & 12.39 & 0.820  & 5000 & 3.10 & +0.22 & 1.10 \\
 114 & 12.07 & 0.841  & 4940 & 2.74 & +0.11 & 1.14 \\    
 129 & 12.25 & 0.838  & 5000 & 2.98 & +0.05 & 1.11 \\
 157 & 12.25 & 0.820  & 5050 & 2.92 & +0.16 & 1.12 \\
                \noalign{\smallskip}
            \hline
         \end{tabular}
      \]
References: $V$ and $b-y$ from Bruntt et al.\ (1999);

$^*$ Star numbers from Lindoff (1972) 
   \end{table}

\input epsf
\begin{figure}
\epsfxsize=\hsize 
\epsfbox[-5 -5 600 430]{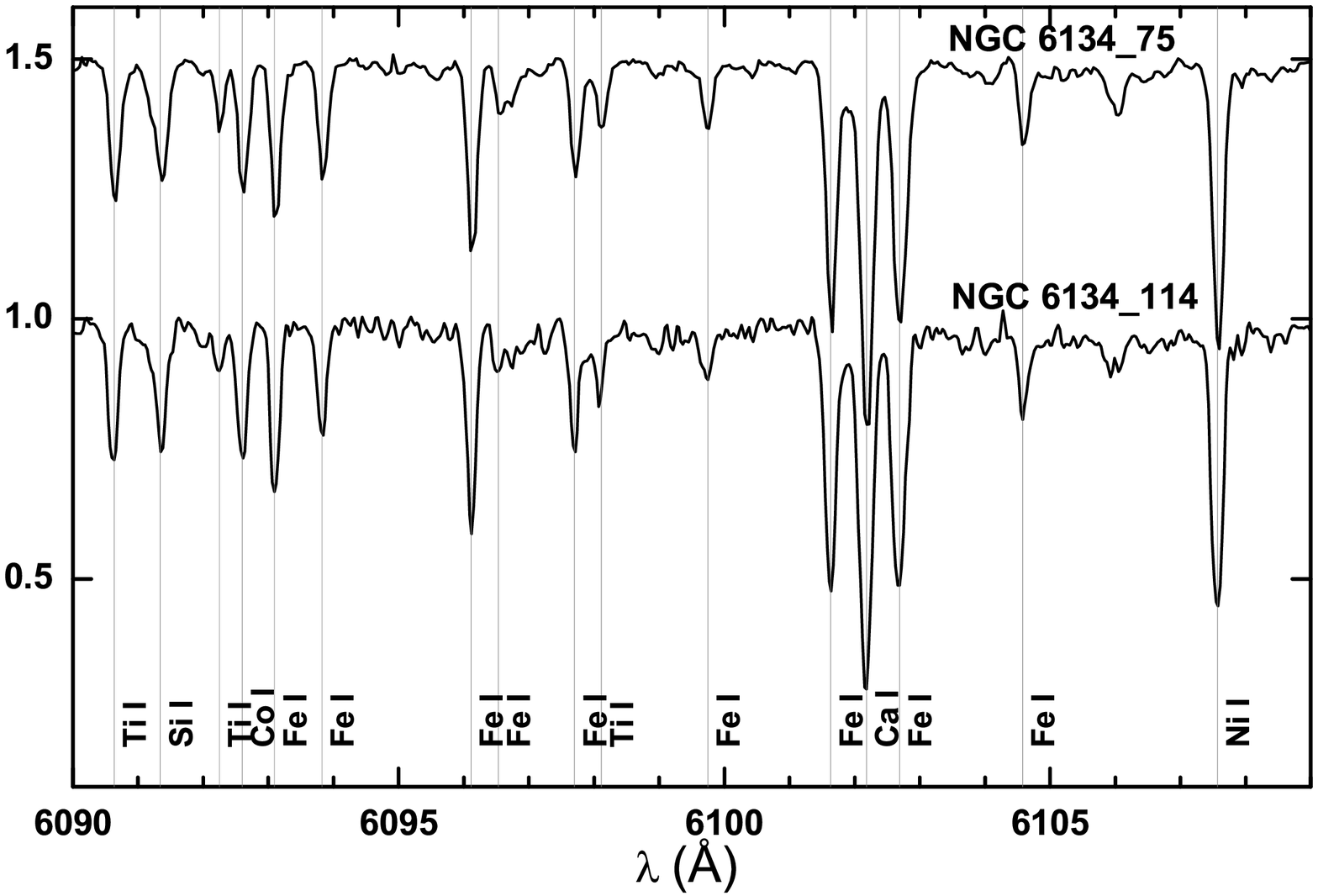} 
    \caption {Small samples of stellar spectra of programme stars in NGC\,6134. Spectra obtained
    with UVES (NGC\,6134\_75)
and FEROS (NGC\,6134\_114) are shown. An offset of 0.5 in relative flux 
is applied for clarity.}
    \label{CMD}
  \end{figure}

In this work we used the same model atmospheres and computing codes of the other BOCCE
program investigations (see Carretta et al. 2004).
Definition of the continuum for the determination of the EWs
is critical for program stars. These objects are cool, of low gravity and high metallicity stars. The spectra
 were normalized to the continuum using ROSA software (Gratton 1988), visually checking the output. 
The lines suitable for measurement were chosen using the requirement that the profiles be 
sufficiently clean to provide reliable equivalent widths. Inspection of the 
solar spectrum (Kurucz et al.\ 1984) and the solar line identifications of 
Moore et al.\ (1966) were used to avoid blends. Lines blended by 
telluric absorption lines were omitted from treatment as well. 
In order to avoid NLTE overionization effects, mainly weak lines were selected 
for the analysis. Abundances of Na and Mg were determined with NLTE taken into 
account as described by Gratton et al.\ (1999). 

The determination of carbon, nitrogen, oxygen, zirconium, yttrium, barium, lanthanum,
cerium, neodymium, and europium abundances were performed using spectral synthesis. 

For  ${\rm C}_2$ determination in stars observed with FEROS we used 5632 -- 5636~{\AA} interval to compare 
with observations of ${\rm C}_2$ Swan (0,1) band head at 5630.5~{\AA}. The same 
molecular data of ${\rm C}_2$ as used by Gonzalez et al.\ (1998) were adopted for the analysis.
For the stars observed with UVES this spectral interval was not available, so we analysed several other Swan (1,0)
bands at 4732.8~{\AA} and 4735.3~{\AA} with the molecular input data from Kurucz \& Bell (1995). 
The interval 7980 -- 8130~{\AA} contains strong $^{12}{\rm C}^{14}{\rm N}$ and $^{13}{\rm C}^{14}{\rm N}$ 
features, so it was used for nitrogen abundance and $^{12}{\rm C}/^{13}{\rm C}$ ratio analysis. 
The molecular data for this CN band were provided by Bertrand Plez (University of Montpellier II). 
All $gf$ values were increased by +0.021~dex to fit the model spectrum of solar atlas of Kurucz et. al.\ (1984).
We derived oxygen abundance from synthesis of the forbidden [O\,{\sc i}] line at 6300~{\AA}. 
The $gf$ values for $^{58}{\rm Ni}$ and $^{60}{\rm Ni}$ isotopic line components, which blend the 
oxygen line, were taken from Johansson et al.\ (2003). 

   \begin{table}
      \caption{Effects on derived abundances resulting from model changes 
for the star NGC\,6134\_114. The table entries show the effects on the 
logarithmic abundances relative to hydrogen, $\Delta$[A/H]. Note that the 
effects on ``relative" abundances,  [A/Fe], are often 
considerably smaller than abundances relative to hydrogen, [A/H] } 
      \[
         \begin{tabular}{lrrc}
            \hline
	    \hline
            \noalign{\smallskip}
Species & ${ \Delta T_{\rm eff} }\atop{ +100 {\rm~K} }$ & 
            ${ \Delta \log g }\atop{ +0.3 }$ & 
            ${ \Delta v_{\rm t} }\atop{ +0.3 {\rm km~s}^{-1}}$ \\ 
            \noalign{\smallskip}
            \hline
            \noalign{\smallskip}
   C\,(C$_2$)  &--0.05 &  0.03 & --0.01 \\
   N\,(CN)    &  0.05 &  0.01 &  0.04  \\
   O\,([O{\sc i}])    &--0.02 &--0.05 &--0.01  \\  
   Na\,{\sc i}   &  0.08 &--0.08 &--0.11  \\  
   Mg\,{\sc i}   &  0.05 &--0.03 &--0.08  \\
   Al\,{\sc i}   &  0.07 &--0.01 &--0.05  \\  
   Si\,{\sc i}   &--0.01 &  0.04 &--0.04  \\    
   Ca\,{\sc i}   &  0.10 &--0.01 &--0.07  \\
   Sc\,{\sc ii}  &--0.01 &  0.14 &--0.08  \\  
   Ti\,{\sc i}   &  0.14 &--0.01 &--0.06  \\    
   Ti\,{\sc ii}  &  0.10 &--0.03 &--0.08  \\       
   V\,{\sc i}    &  0.16 &  0.00 &--0.06  \\  
   Cr\,{\sc i}   &  0.09 &--0.01 &--0.09  \\
   Cr\,{\sc ii}  &  0.10 &--0.03 &--0.07  \\
   Mn\,{\sc i}   &  0.08 &--0.02 &--0.05  \\
   Co\,{\sc i}   &--0.09 &  0.03 &--0.05  \\
   Ni\,{\sc i}   &--0.05 &  0.03 &--0.12  \\
   Cu\,{\sc i}   &  0.01 &  0.02 &  0.01  \\
   Zn\,{\sc i}   &--0.02 &  0.04 &--0.09  \\
   Y\,{\sc ii}    &--0.02 &  0.11 &--0.13  \\
   Zr\,{\sc i}    &--0.18 &  0.00 &--0.03  \\
   Ba\,{\sc ii}   &--0.10 &  0.09 &--0.10  \\
   La\,{\sc ii}   &  0.01 &--0.01 &  0.04  \\
   Ce\,{\sc ii}  &  0.00 &  0.13 &--0.01  \\
   Nd\,{\sc ii}  &--0.02 &  0.11 &--0.01  \\
   Eu\,{\sc ii}  &  0.00 &  0.10 &--0.01  \\
 $^{12}{\rm C}/^{13}{\rm C}$ & -1 & -1 & 0 \\
   
                 \noalign{\smallskip}
            \hline
         \end{tabular}
      \]
   \end{table}


Zirconium abundance was derived using Zr\,{\sc i} lines at 4687~{\AA} and 6127~{\AA}, barium 
from Ba\,{\sc ii} 5853~{\AA} and 6496~{\AA},  lanthanum from La\,{\sc ii} lines at 6320~{\AA} and 6390~{\AA}, 
cerium from 
Ce\,{\sc ii} lines at 5274~{\AA} and 6043~{\AA}, neodymium from Nd\,{\sc ii} 5093~{\AA}, 5293~{\AA} 
and 5320~{\AA} lines. Europium abundance was derived using Eu\,{\sc ii} line at 6645~{\AA}. 
The hyperfine structure for Eu\,{\sc ii} line was used for the synthesis. 
The wavelength, excitation energy and total log~$gf = 0.12$ were taken from Lawler et al.\ (2001), the
isotopic fractions of $^{151}{\rm Eu}$ 47.77\% and $^{153}{\rm Eu}$ 52.23\%, and isotopic
shifts were taken from Biehl (1976).

Oscillator strengths for the most important lines of other elements were taken mainly from an inverse solar 
spectrum analysis done in Kiev (Gurtovenko \& Kostik 1989). 

\subsection{Estimation of uncertainties}

The sources of uncertainty can be divided into two categories. The first 
category includes  the errors which affect all 
the lines together, i.e.\ mainly the model errors (such as errors in the 
effective temperature, surface gravity, microturbulent velocity, etc.). 
The second category 
includes the errors that affect a single line (e.g.\ random errors 
in equivalent widths, oscillator strengths), i.e.\ uncertainties of the 
line parameters.

  
Typical internal error estimates for the atmospheric parameters are: 
$\pm~100$~K for $T_{\rm eff}$, $\pm 0.3$~dex for log~$g$ and 
$\pm 0.3~{\rm km~s}^{-1}$ for $v_{\rm t}$. The sensitivity of the abundance 
estimates to changes in the atmospheric parameters by the assumed errors is 
illustrated  for the star NGC\,6134\_114 (Table~2). Possible 
parameter errors do not affect the abundances seriously; the element-to-iron 
ratios, which we use in our discussion, are even less sensitive. 

The sensitivity of iron abundances to stellar atmospheric parameters were described 
in Carretta et al.\ (2004). The changes in temperature of 90~K, log~$g$ of 0.1~dex,
and 0.1~km\,s$^{-1}$ for microturbulent velocity
lead to $\Delta{\rm [Fe/H]_I}=0.057$,
 $\Delta{\rm [Fe/H]_I}=0.011$, $\Delta{\rm [Fe/H]_I}=-0.044$ and 
 $\Delta{\rm [Fe/H]_{II}}=-0.086$,
 $\Delta{\rm [Fe/H]_{II}}=0.107$, $\Delta{\rm [Fe/H]_{II}}=-0.043$, respectively. 

Since abundances of C, N and O are bound together by the molecular equilibrium 
in the stellar atmosphere, we have also investigated how an error in one of 
them typically affects the abundance determination of another. 
$\Delta{\rm [O/H]}=0.10$ causes 
$\Delta{\rm [C/H]}=0.05$ and $\Delta{\rm [N/H]}=-0.10$;  
$\Delta{\rm [C/H]}=0.10$ causes $\Delta{\rm [N/H]}=-0.15$ and 
$\Delta{\rm [O/H]}=0.05$. $\Delta {\rm [N/H]}=0.15$ has no effect
on either the carbon or the oxygen abundances.

The scatter of the deduced line abundances $\sigma$, presented in Table~3, 
gives an estimate of the uncertainty due to the random errors, e.g. in 
continuum placement and the line parameters (the mean value of  $\sigma$ 
is $0.06$ dex).
 Thus the uncertainties in the derived abundances that are the 
result of random errors amount to approximately this value. 

\section{Results and discussion}

The abundances relative to hydrogen
[El/H]\footnote{In this paper we use the customary spectroscopic notation
[X/Y]$\equiv \log_{10}(N_{\rm X}/N_{\rm Y})_{\rm star} -
\log_{10}(N_{\rm X}/N_{\rm Y})_\odot$} and $\sigma$ (the line-to-line 
scatter) derived for up to 26 neutral and ionized species for the programme 
stars are listed in Table~3.
The average cluster abundances and dispersions about the mean values for NGC\,6134 are 
presented in Table~3 as well. 

\subsection{Galactic radial abundance gradient}

The open cluster radial abundance gradient was analysed and discussed 
many times during several decades (see e.g. Friel 1995, Twarog et al.\ 1997; Bragaglia et al.\ 2001, 2008; 
Friel et al.\ 2002, 2005; Carretta et al.\ 2004, 2005, 2007; Yong et al.\ 2005; Sestito et al.\ 2006, 2007, 2008; 
Jacobson et al.\ 2008, 2009; Smiljanic et al.\ 2009; Pancino et al.\ 2010 and references therein).
Twarog et al.\ (1997) first proposed that the open cluster abundance distribution is not a negative linear gradient 
but two separate distributions, each of constant metallicity, divided at $R_{\rm gc}= 10$ kpc. 
Clusters in the inner part are of solar metallicity, while in the outer part the mean metallicity is about 
$-0.3$~dex. The recent investigations of open clusters, reaching also the more distant parts of the disk, show that 
maybe the gradient is not the same in all the disk: it is rather steep in the inner disk ($R_{\rm gc}< 12-14$~kpc), 
and it flattens in the outer disk (c.f., Carraro et al.\ 2004, 2007; Yong et al.\ 2005; Sestito et al.\ 
2006, 2008; Jacobson et al.\ 2009).

\input epsf
\begin{figure*}
\epsfxsize=\hsize 
\epsfbox[0 0 600 360]{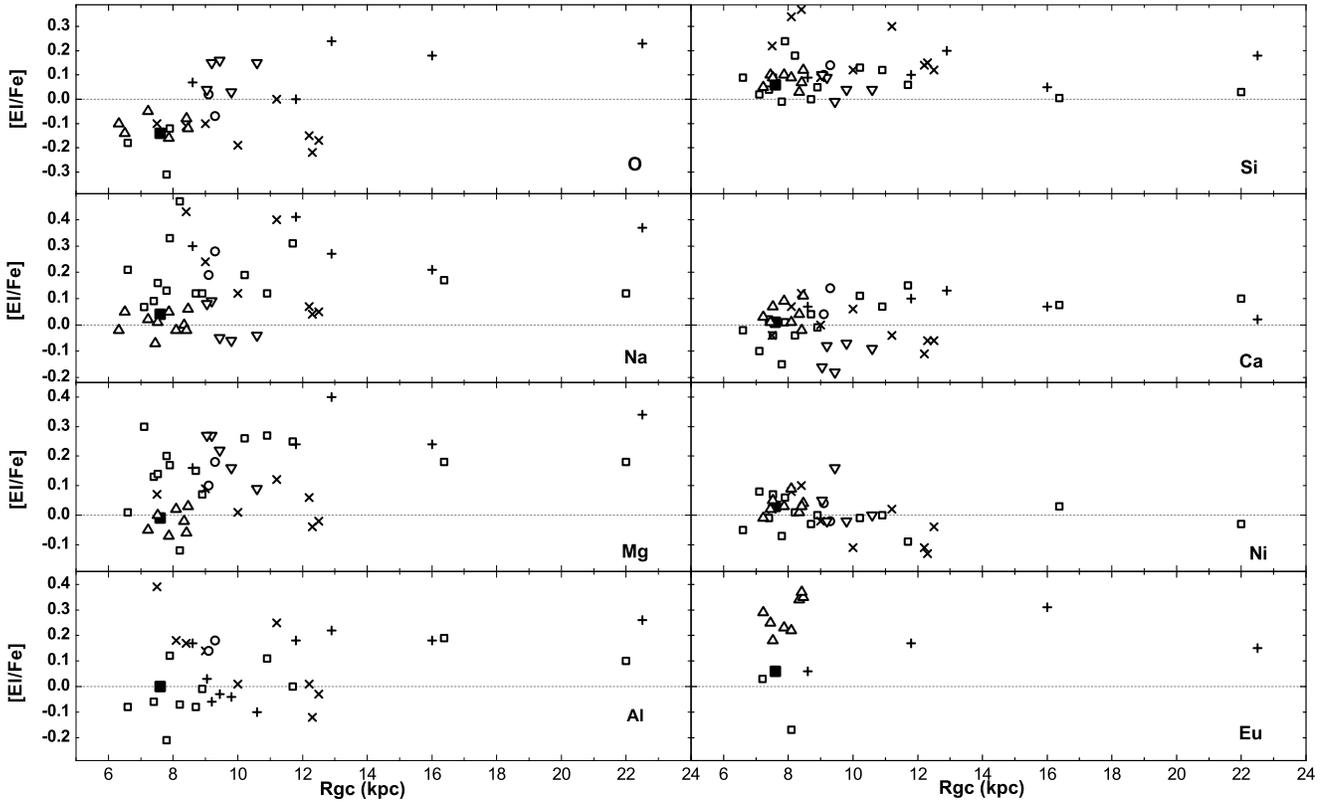} 
\caption {Radial distribution of open cluster abundances. 
Open squares show results by Bragaglia et al.\ (2001, 2008), 
Carretta et al.\ (2004, 2005, 2007), and Sestito et al.\ (2006, 2007, 2008). 
The result of this study is marked as a filled square. 
Crosses show results presented in the papers by Friel et al.\ (2005) and 
Jacobson et al.\ (2008, 2009),  
pluses -- Yong et al.\ (2005),
triangles -- Smiljanic et al.\ (2009), 
and reversed triangles -- Pancino et al.\ (2010).
}
\label{rad.distr.}
\end{figure*}
%

Recently, Magrini et al.\ (2009)  used a set of literature abundances of open
clusters based on
high resolution spectroscopy to compare the gradient, and its time evolution,
with their chemical evolution models.  However, their sample
is not homogeneous (distances, ages, and abundances were taken from papers of
many different groups) so it is not an ideal set, since systematics can mask
or produce features in the distribution. The BOCCE project aims at
collecting a homogeneous sample, 
well spread over the Galactic plane, covering all ages, and metallicities.
Fig.~4 displays the radial distribution of some elemental abundances for 
BOCCE clusters analysed so far, and for others in recent studies. 
The scatter is quite large at all radii, but NGC\,6134 agrees well with results 
of other open clusters at the same $R_{\rm gc}$ of 7.5 kpc.

\subsection{Carbon, nitrogen and oxygen}

Since the [C\,{\sc i}] 8727~{\AA} line is blended with CN in spectra of red
giants, and since the probability of increased  strengths of CN molecular lines
is present, the ${\rm C}_2$ Swan (0,1) band head  at 5635.5~{\AA} is a quite
popular feature for the carbon abundance determinations. It  was used in our
previous studies of giants (e.g. Tautvai\v{s}ien\.{e} et al.\ 2000, 2005). In
Fig.~5, a fit to the NGC\,6134\_157 spectrum at ${\rm C}_2$ 5635.5~{\AA} is
shown.  The Swan band (1,0) with the head at 4737~{\AA} is rarely used for
carbon abundance determinations,  since the region is quite crowded by
additional lines. Fortunately, two features of ${\rm C}_2$ at  4732.8~{\AA} and
4735.3~{\AA} were clearly seen in the spectra obtained with the UVES
spectrograph.  In Fig.~6, we show an example of spectral fit to these ${\rm
C}_2$ bands for the star NGC\,6134\_75. 

\input epsf
\begin{figure}
\epsfxsize=\hsize 
\epsfbox[-20 -20 620 470]{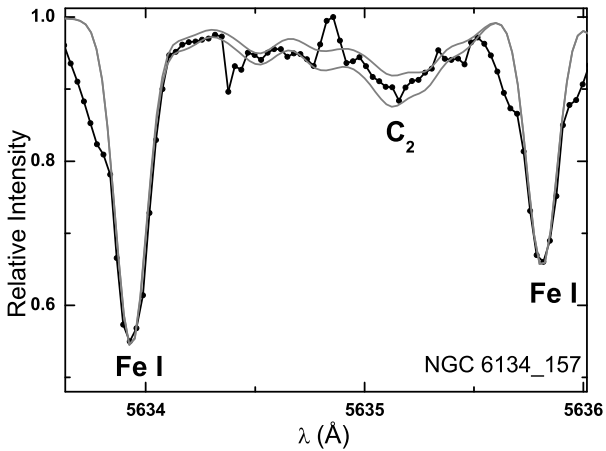} 
    \caption {Small region of NGC\,6134\_157 spectrum (solid black line with black dots) at 
${\rm C}_2$ Swan (0,1) band head 5635.5~{\AA}, plotted together with 
synthetic spectra with [C/Fe] values lowered by $-0.2$~dex (lower gray line) and $-0.25$~dex 
(upper gray line). 
}
    \label{}
  \end{figure}

The interpretation of the carbon abundance in NGC\,6134 can be done by a 
comparison with carbon abundances determined for dwarf stars 
in the Galactic disk. Shi et al.\ (2002) performed an abundance 
analysis of carbon for a sample of 90 F and G type main-sequence disk stars  
using C\,{\sc i} and [C\,{\sc i}] lines and found [C/Fe] to be about solar 
at the solar metallicity.   
The same result was found by Gustafsson et al.\ (1999) 
who analysed a sample of 80 late F and early G type dwarfs using the forbidden 
[C\,{\sc i}] line. 
The ratios of [C/Fe] in our stars lie about 0.2~dex below 
the trend obtained for dwarf stars of the Galactic disk.      
Smiljanic et al.\ (2009) analysed carbon abundances in two stars of the same cluster 
NGC\,6134\_30 and NGC\,6134\_202 and also found [C/Fe] abundance ratios to be 
lowered by about the same amount. 

The interval 7980--8130~{\AA}, with 11 CN lines selected, was analysed  in order
to determine the nitrogen abundances. The mean nitrogen to iron abundance ratio 
in NGC\,6134 is [N/Fe]=$0.25\pm0.10$. Nitrogen to iron abundance ratios in the
two clump  stars investigated in NGC\,6134 by Smiljanic et al.\ (2009)[N/Fe]
ratios are enhanced  by 0.42 and 0.36~dex as well. 

This shows that nitrogen is overabundant in these clump stars of NGC\,6134,
while
[N/Fe]  values in the main-sequence stars are about solar at the solar
metallicity  (c.f. Shi et al.\ 2002). Reddy et al.\ (2003) investigated nitrogen
abundances in a sample  of 43 F--G dwarfs in the Galactic disk by means of weak
N\,{\sc i} lines. At a value of  [Fe/H] of about $-0.2$~dex, which was well
represented in their sample, [N/Fe] is about 0.2~dex.  There were few stars of
solar metallicity investigated in this study. Nevertheless, the authors  make
the extrapolation that at solar metallicity [N/Fe] values should be solar. 

C/N ratios in our programme stars of NGC\,6134 range from $0.98$ to $1.48$. For the stars 
NGC\,6134\_30 and NGC\,6134\_202, Smiljanic et al.\ also found C/N to be $0.99$ 
and $1.41$, respectively. 

\input epsf
\begin{figure*}
\epsfxsize=\hsize 
\epsfbox[-20 -20 650 260]{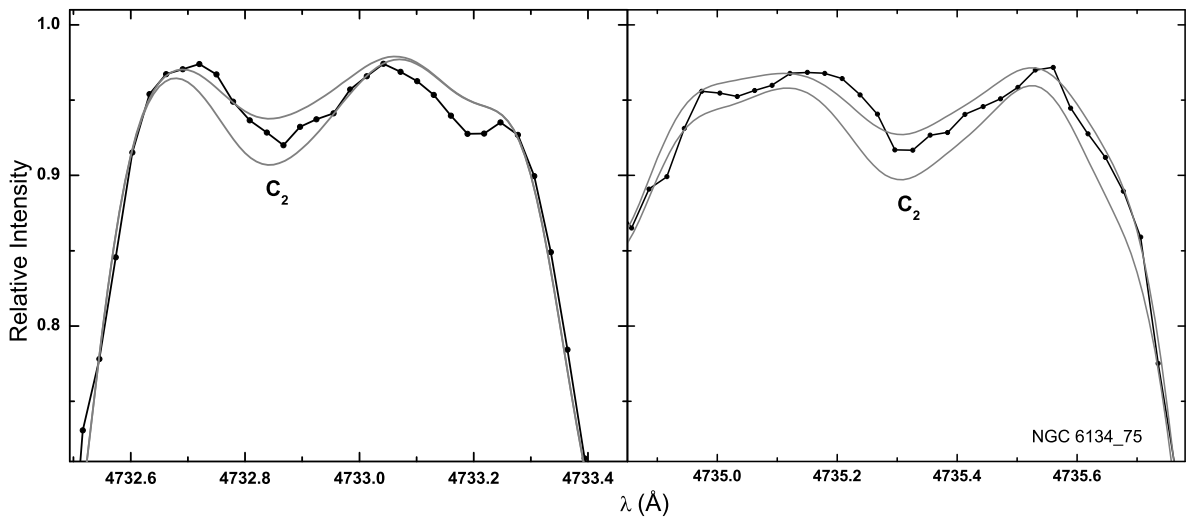} 
    \caption {Small region of NGC\,6134\_75 spectrum (solid black line with black dots) at 
${\rm C}_2$ Swan (1,0) band heads 4732.8~{\AA} and 4735.3~{\AA}, plotted together with 
synthetic spectra with [C/Fe] values lowered by $-0.2$~dex (lower gray line) and $-0.25$~dex 
(upper gray line). 
}
    \label{}
  \end{figure*} 

The solar carbon and nitrogen abundances used in our work are log$A_{\rm C} = 8.52$ and 
log$A_{\rm N} = 7.92$ (Grevesse \& Sauval 2000), so the solar ${\rm C/N} = 3.98$.
Smiljanic et al.\ (2009) presented carbon and nitrogen values for ten open clusters in total 
and found an average C/N ratio of about $0.98$.

\input epsf
\begin{figure}
\epsfxsize=\hsize 
\epsfbox[-20 -20 620 500]{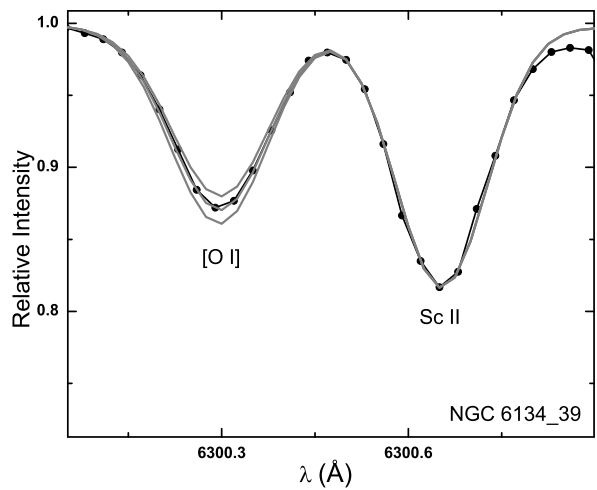} 
    \caption {Fit to the forbidden [O\,{\sc i}] line at 6300~{\AA} in 
NGC\,6134\_39. The observed spectrum is shown as a solid line with black dots. Synthetic
spectra with [O/Fe]$=-0.3$, $-0.25$, and $-0.2$ are shown as solid gray lines.}
    \label{Oxygen}
  \end{figure}

The analysis of oxygen abundance was performed using the most popular forbidden [O\,{\sc i}] line 
at 6300~{\AA}. In the spectra of NGC\, 6134 stars this line is not contaminated by telluric lines. 
In Fig.~7 we show an example of spectrum syntheses for 
[O\,{\sc i}] line in NGC\,6134\_39 star. This line is blended by Ni\,{\sc i}. The atomic data 
of Ni\,{\sc i} were taken from Johansson et al.\ (2003). 
   
The oxygen abundance results as a function of galactocentric distance are shown in Fig.~4. 
Similarly to other open clusters, [O/Fe] in NGC\,6134 stars is lower than the solar value by 
about 0.2~dex. Oxygen abundances were not investigated in NGC\,6134 by Smiljanic et al.\ (2009). 

In Fig.~8, [O/Fe] are plotted as a function of C/N ratios. 
The values of NGC\,6134 are in agreement with other studies. At smaller C/N ratios [O/Fe] are 
lower. 

\input epsf
\begin{figure}
\epsfxsize=\hsize 
\epsfbox[-20 -20 620 500]{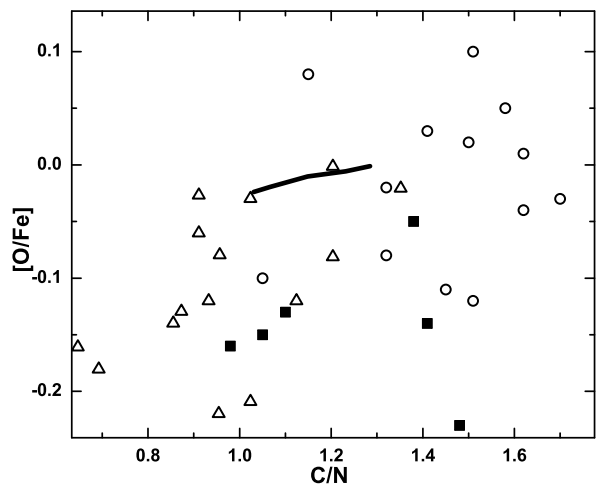} 
    \caption { The oxygen abundance as a function of carbon and nitrogen abundance ratio. NGC\,6134   
    stars are shown as filled squares, stars from open clusters M~67 and NGC\,7789 investigated by Tautvai\v{s}ien\.e et 
    al.\ (2000, 2005) as circles, results by Smiljanic et al.\ (2009) -- triangles. The solid line shows the values 
predicted by Schaller et al.\ (1992).}
    \label{Oxygen}
  \end{figure}

\subsection{Carbon isotope ratios}

The $^{12}{\rm C}/^{13}{\rm C}$ ratios were determined for all programme stars from the (2,0)
$^{13}{\rm C}^{14}{\rm N}$ feature at 8004.728~{\AA} with a laboratory wavelength
adopted from Wyller (1966). In Fig.~9 we show a small portion of NGC\,6134\_157 spectrum together with three 
theoretical lines for carbon isotopic ratio determination.
We find that the mean $^{12}{\rm C}/^{13}{\rm C}$ ratios are lowered 
to about  $9\pm2.5$ in the clump stars investigated. 
Smiljanic et al.\ (2009) found for this ratio the value of 12 and 13 for their
two stars in NGC 6124. The $^{12}{\rm C}/^{13}{\rm C}$ ratio in the solar photosphere is equal to 89 
(Coplen et al.\ 2002).  
 
The number of papers with carbon isotope ratios determined for red clump stars in open 
clusters is not numerous. 
In Fig.~10 we plot the mean carbon isotope ratios of clump stars in different open clusters as a function of 
turn-off mass and compare them
with the theoretical models of the $1^{st}$ dredge-up by Boothroyd \& Sackman (1999) 
and Charbonnel (1994). 

For NGC\,6134 we plot the mean $^{12}{\rm C}/^{13}{\rm C}$ ratio $10\pm 3$ as 
averaged from clump stars investigated in our study and by Smilanic et al.\ (2009).
The mass of the programme stars $M = 2.34M_{\odot}$ was obtained by Carretta et al.\ (2004) reading the 
turn-off values on the Girardi et al.\ (2000) isochrones for solar metallicity at the age
of the cluster of 0.7~Gyr as determined by Bruntt et al.\ (1999). 

In Fig.~10, we also present results for clump stars in 5 other open clusters investigated by  
Smiljanic et al.\ (2009). Data for M\,67 and NGC\,7789 clump stars come from Tautvai\v{s}ien\.{e} 
et al.\ (2000 and 2005). From Gilroy (1989) we selected 4 clusters with well defined 
red clump stars (NGC\,752, NGC\,2360, M\,67, and IC\,4756). Luck (1994) derived carbon isotope ratios for 8 open 
clusters, however it is very difficult to identify stars of red clump in them.
 
From Fig.~10, it is seen that $^{12}{\rm C}/^{13}{\rm C}$ values in the clump stars are much below the predictions 
of $1^{st}$ dredge-up, and for most of the clusters they
are even below the Cool Bottom Processing (CBP) model of 
extra-mixing (Boothroyd \& Sackman 1999). 
 
\input epsf
\begin{figure}
\epsfxsize=\hsize 
\epsfbox[0 0 620 500]{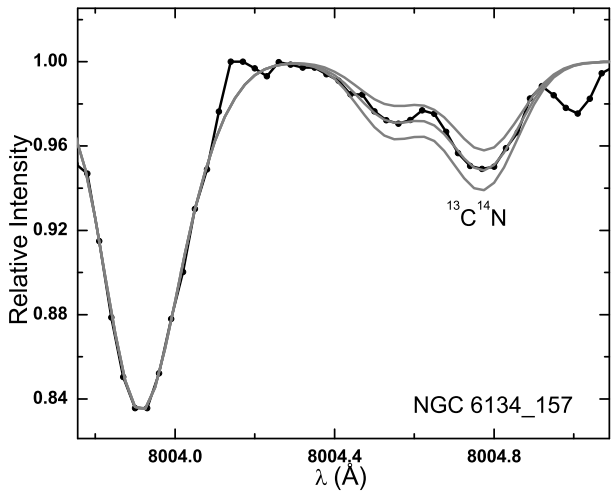} 
    \caption {Small region of NGC\,6134\_157 spectrum (solid black line with black dots) 
    with $^{13}{\rm C}^{14}{\rm N}$ feature.
Grey lines show synthetic spectra with $^{12}{\rm C}$/$^{13}{\rm C}$ ratios equal to 10 (lower line), 
12 (middle line) and 14(upper line). 
}
    \label{Oxygen}
  \end{figure}


\input epsf
\begin{figure}
\epsfxsize=\hsize 
\epsfbox[-20 -20 620 500] {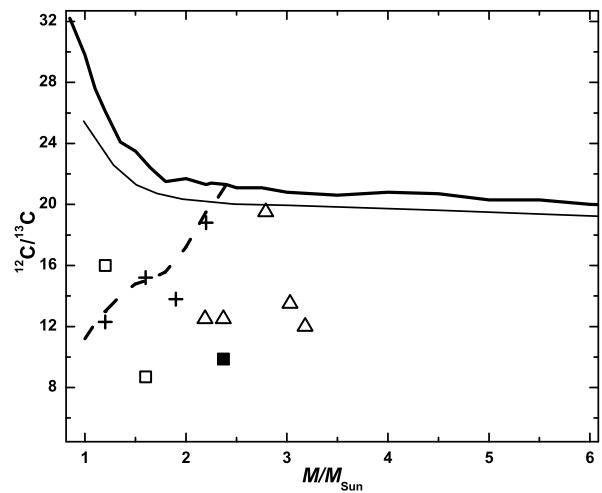} 
    \caption {The averaged carbon isotope ratios in clump stars of open clusters
as a function of stellar  turn-off mass. $^{12}{\rm C}$/$^{13}{\rm C}$ in
NGC\,6134 (as averaged from this work and Smiljanic et al.)  is shown as a
filled square, in M~67 and NGC\,7789 as open squares (Tautvai\v{s}ien\.{e} et
al.\ 2000, 2005). Data for clusters investigated by Smiljanic et al.\ (2009)
are shown as triangles, and by Gilroy (1989) as plus  signs. The thick solid
curve presents the theoretical model of $1^{st}$ dredge-up for  $Z=0.02$ by
Boothroyd \& Sackman (1999) and the thin solid line the
one by Charbonnel (1994). The
CBP model of  extra-mixing by Boothroyd \& Sackman (1999) is plotted as a thick
dashed line. 
}
    \label{Oxygen}
  \end{figure}

\subsection{Sodium and Aluminum}

Sodium and aluminum are among the chemical elements for which observations of 
abundance anomalies are also present. The O-Na anticorrelation has been observed 
among the brightest red giants in Galactic globular clusters for a long time 
(see Kraft 1994; Da Costa 1998; Denissenkov \& Herwig 2003 and references therein).
An overabundance of Na could appear, due to the deep mixing from layers of 
the NeNa cycle of H burning. Extensive theoretical studies of deep mixing in 
stellar atmospheres have been made by Denissenkov \& Weiss (1996), 
Denissenkov \& Tout (2000), Denissenkov \& Herwig (2003), Gratton et al.\ (2004) and references therein.   
However, the explanation of abundance changes by deep mixing has been  
eliminated by the determination of an Na--O anticorrelation in less evolved stars 
down to the main sequence (Gratton et al.\ 2001; Thevenin et al.\ 2001; 
Ramirez \& Cohen 2002; D'Orazi et al. 2010). For a recent discussion of the Na
and O abundances in open clusters, see De Silva et al. (2009), who explicitly
addressed the problem of the (not seen) Na--O anticorrelation.
 
Abundances of sodium were determined from the NLTE analysis of Na\,{\sc i} lines 
at 5682.64~{\AA}, 6154.23~{\AA} and 6160.75~(\AA). 
The Na\,{\sc i} line at 5682.64~{\AA} was not available for 
the analysis of UVES spectra.
Abundances of aluminum were determined from Al\,{\sc i} lines at 7835.30~\AA\  and 7836.13~{\AA}. 

The stars in our sample do not show
overabundance either of sodium, or of aluminum (Fig.~4).

\subsection{The $\alpha$-elements, iron group and heavier elements}

According to observations of main sequence stars in the Galactic disk, 
abundance ratios of $\alpha$-process elements to iron at the solar 
metallicity are solar or slightly higher. In the study by Edvardsson et al.\ 
(1993) [Mg/Fe], [Si/Fe] and [Ti/Fe] for almost all of the stars lie slightly 
above the solar ratio, [Ca/Fe] are solar. In the study of 
Reddy et al.\ (2003), [Mg/Fe] and [Si/Fe] values are above solar, while 
[Ca/Fe] and [Ti/Fe] are exactly solar. 
The mean [$\alpha$/Fe] ratios in a majority of open clusters investigated 
are slightly higher than in the Sun, and especially the overabundance 
of silicon is noticeable 
(c.f. Brown et al.\ 1996; Bragaglia et al.\ 2001; Friel et al.\ 2003; but see
also
Bragaglia et al. 2008, Sestito et al. 2008). There has also been a claim that
[$\alpha$/Fe] is higher for outer disk clusters (e.g., Yong et al. 2005), but
this has not been confirmed by other studies (Carraro et al. 2007, Sestito et
al. 2008).

In NGC\,6134, the mean cluster [$\alpha/{\rm Fe}] \equiv {1\over 4}
([{\rm Mg}/{\rm Fe}]+[{\rm Si}/{\rm Fe}]+[{\rm Ca}/{\rm Fe}]
+[{\rm Ti}/{\rm Fe}]) = 0.02\pm0.03$~(s.d.), which is very close to the solar value. 

As it is seen from Table~3, the ratios of abundances of iron 
group, $s-$ ,and $r-$process elements to iron 
are close to solar.

\subsection{Final remarks}

Convection, the only mechanism of internal mixing taken into account by 
standard stellar evolution models, is not able to account for carbon and 
nitrogen abundance alterations seen in clump stars of open clusters. 
The clump stars provide information on 
chemical composition changes, which have happened during their evolution along 
the giant branch and during the helium flash.
Extra-mixing processes may become efficient on the red giant branch when 
stars reach the so-called RGB  bump, and may modify the surface 
abundances. It is known that alterations of $^{12}{\rm C}/^{13}{\rm C}$ and 
$^{12}{\rm C}/^{14}{\rm N}$ ratios depend on stellar evolutionary stage, mass and 
metallicity (see Charbonnel et al.\ 1998, Gratton et al.\ (2000), Chanam\'{e} et al.\ 2005, 
Smiljanic et al.\ 2009 for more discussion). 
 
In the open cluster M\,67 (Tautvai\v{s}ien\.{e} et al.\ 2000), we did not find 
evidence for extra-mixing in first-ascent giants (the mass of turn-off stars in 
this cluster is about 1.2~$M_{\odot}$).  
In the M\,67 giants investigated, the mean $^{12}{\rm C}/^{13}{\rm C}$ ratio is 
lowered to the value of $24 \pm 4$ and the $^{12}{\rm C}/^{14}{\rm N}$ ratio to 
the value of $1.7 \pm 0.2$, which is close to the corresponding predictions of the 
first dredge-up (Boothroyd \& Sackmann 1999). Evidence of extra-mixing has been  
detected only in the clump stars observed, where
the mean $^{12}{\rm C}/^{13}{\rm C}$ ratio is 
lowered to the value of $16 \pm 4$, and the  
$^{12}{\rm C}/^{14}{\rm N}$ ratio to  $1.4 \pm 0.2$. In  giants 
and clump stars investigated by Tautvai\v{s}ien\.{e} et al.\ (2005) 
in NGC\,7789, $^{12}{\rm C}/^{13}{\rm C}$ 
ratios are about the same, $9\pm 1$, but $^{12}{\rm C}/^{14}{\rm N}$ ratio is
$1.9\pm 0.5$ in giants and $1.3\pm 0.2$ in clump stars. 

In NGC\,6134 we investigated only clump stars. The mean $^{12}{\rm C}/^{14}{\rm N}$ 
ratio in these clump stars, $1.2\pm 0.2$, and the mean $^{12}{\rm C}/^{13}{\rm C}$ ratio 
$9\pm 2.5$, indicate a large extra-mixing, which is even larger than 
it is foreseen by the CBP model (Boothroyd \& Sackmann 1999) 
for stars of a similar turn-off mass. $^{12}{\rm C}/^{13}{\rm C}$ ratios for the majority 
of clump stars in other open clusters also lie below the CBP trend (Gilroy 1989; 
Tautvai\v{s}ien\.{e} et al.\ 2000, 2005; Smiljanic et al.\ 2009).
Further studies are required both on the theoretical and observational sides in order 
to improve our understanding of mixing processes in low and intermediate mass giants.

\section*{Acknowledgments}

This research has made use of Simbad, VALD and NASA ADS databases.
Bertrand Plez (University of Montpellier II) and Guillermo Gonzalez  
(Washington State University) were particularly generous in providing us with 
atomic data for CN and C$_2$ molecules, respectively.
\v{S}.\,M. and G.\,T. were supported by the Ministry of Education and Science of Lithuania 
via LitGrid programme and by the European Commission via FP7 Baltic Grid II project. 


\newpage
\begin{landscape}
\begin{table} \scriptsize
\centering

  \caption{Abundances relative to hydrogen [El/H]. The quoted 
errors, $\sigma$, are the standard deviations in the mean value due to the 
line-to-line scatter within the species. The number of lines used is indicated by $n$. 
The last two columns give the mean [El/Fe] and standard deviations for the cluster stars 
investigated.}

\begin{tabular}{lrrrrrrrrrrrrrrrrrrrrrrrrrr}
  \hline
  \hline
\scriptsize
  & \multicolumn{3}{c}{114} &
  & \multicolumn{3}{c}{39} &
  & \multicolumn{3}{c}{157} &
  & \multicolumn{3}{c}{75} &
  & \multicolumn{3}{c}{129} &
  & \multicolumn{3}{c}{69} &
  & \multicolumn{2}{c}{Mean}\\
            \noalign{\smallskip}
\cline{2-4}\cline{6-8}\cline{10-12}\cline{14-16}\cline{18-20}\cline{22-24}\cline{26-27}
            \noalign{\smallskip}
Species &[El/H] &$\sigma$ &$n$&\ &[El/H] &$\sigma$ &$n$&\ &[El/H] &$\sigma$ &$n$&\ &[El/H] &$\sigma$ &$n$&\ &[El/H] &$\sigma$ &$n$&\ &[El/H]&$\sigma$&$n$&\ &[El/Fe] & $\sigma$\\ 
            \noalign{\smallskip}
            \hline
            \noalign{\smallskip}
C\,(C$_2$) &--0.15&  & 1  & &  0.04 &        & 1  & & --0.07 &       &  1  & &--0.05 & 0.04   & 2  & & --0.20  &   0.07 & 2  & &--0.25 & 0.04  & 2  & &--0.26&   0.11   \\
N\,(CN)         & 0.43	& 0.14  & 11 & &  0.47 &  0.16  & 10 & &   0.49 & 0.14  & 11  & &  0.40 & 0.05   & 11 & &   0.41  &   0.07 & 11 & &  0.21 & 0.06  & 11 & & 0.25	&   0.10   \\        
O\,([O{\sc i}]) &--0.04	&       & 1  & &  0.01 &        & 1  & &   0.03 &       &  1  & &  0.08 &        & 1  & & --0.05  &        & 1  & &  0.00 &       & 1  & &--0.14&   0.05   \\
Na\,{\sc i}     & 0.13	& 0.07  & 3  & &  0.17 &  0.06  & 3  & &   0.17 & 0.06  &  3  & &  0.28 & 0.04   & 2  & &   0.25  &   0.07 & 2  & &  0.15 & 0.04  & 2  & & 0.04	&   0.06   \\        
Mg\,{\sc i}  	& 0.07	& 0.06  & 4  & &  0.15 &  0.06  & 4  & &   0.11 & 0.07  &  4  & &  0.23 & 0.04   & 3  & &   0.17  &   0.03 & 3  & &  0.10 & 0.01  & 3  & &--0.01&   0.06   \\
Al\,{\sc i}  	& 0.08	& 0.08  & 2  & &  0.32 &  0.07  & 2  & &   0.28 & 0.06  &  2  & &  0.15 & 0.02   & 2  & &   0.08  &   0.04 & 2  & &  0.00 & 0.04  & 2  & & 0.00	&   0.12   \\        
Si\,{\sc i}   	& 0.20	& 0.08  & 12 & &  0.17 &  0.07  & 12 & &   0.21 & 0.06  &  12 & &  0.36 & 0.09   & 5  & &   0.21  &   0.07 & 5  & &  0.12 & 0.09  & 5  & & 0.06	&   0.08   \\        
Ca\,{\sc i}  	& 0.21	& 0.06  & 12 & &  0.38 &  0.08  & 12 & &   0.20 & 0.05  &  12 & &  0.10 & 0.07   & 9  & &   0.06  &   0.04 & 9  & &  0.00 & 0.07  & 9  & & 0.01	&   0.14   \\        
Sc\,{\sc ii}  	& 0.20	& 0.05  & 9  & &  0.25 &  0.08  & 9  & &   0.20 & 0.08  &  9  & &  0.31 & 0.03   & 3  & &   0.13  &   0.03 & 3  & &  0.05 & 0.05  & 3  & & 0.04	&   0.09   \\        
Ti\,{\sc i}  	& 0.25	& 0.07  & 18 & &  0.27 &  0.08  & 18 & &   0.23 & 0.03  &  18 & &  0.18 & 0.08   & 5  & &   0.05  &   0.06 & 5  & &--0.01 & 0.09  & 5  & & 0.01	&   0.12   \\        
Ti\,{\sc ii}  	& 0.10	& 0.05  & 8  & &  0.22 &  0.06  & 8  & &   0.04 & 0.04  &  8  & &  0.30 & 0.09   & 2  & &   0.40  &   0.07 & 2  & &  0.30 & 0.01  & 2  & & 0.08	&   0.14   \\        
V\,{\sc i}  	& 0.23	& 0.03  & 9  & &  0.39 &  0.04  & 9  & &   0.24 & 0.05  &  9  & &  0.32 & 0.09   & 2  & &   0.22  &   0.08 & 2  & &  0.10 & 0.07  & 2  & & 0.10	&   0.10   \\        
Cr\,{\sc i}  	& 0.08	& 0.08  & 24 & &  0.25 &  0.08  & 24 & &   0.11 & 0.07  &  24 & &  0.30 & 0.08   & 12 & &   0.27  &   0.05 & 12 & &  0.15 & 0.09  & 12 & & 0.05	&   0.09   \\  
Cr\,{\sc ii}  	& 0.03	& 0.06  & 6  & &  0.24 &  0.03  & 6  & &   0.10 & 0.06  &  6  & &  0.25 &        & 1  & &   0.25  &        & 1  & &  0.17 &       & 1  & & 0.03	&   0.09   \\  
Mn\,{\sc i}  	& 0.28	& 0.03  & 6  & &  0.32 &  0.08  & 6  & &   0.32 & 0.05  &  6  & &  0.25 & 0.08   & 2  & &   0.15  &   0.04 & 2  & &  0.05 & 0.07  & 2  & & 0.08	&   0.11   \\  
Co\,{\sc i}   	& 0.24	& 0.07  & 8  & &  0.35 &  0.06  & 8  & &   0.40 & 0.05  &  8  & &  0.31 & 0.07   & 4  & &   0.18  &   0.02 & 4  & &  0.08 & 0.02  & 4  & & 0.11	&   0.12   \\  
Ni\,{\sc i}   	& 0.17	& 0.08  & 36 & &  0.27 &  0.07  & 36 & &   0.31 & 0.07  &  36 & &  0.24 & 0.07   & 21 & &   0.11  &   0.07 & 21 & &--0.01 & 0.08  & 21 & & 0.03	&   0.12   \\  
Cu\,{\sc i}   	& 0.13	& 0.06  & 3  & &  0.24 &  0.07  & 3  & &   0.27 & 0.05  &  3  & &  0.32 &        & 1  & &   0.30  &        & 1  & &  0.10 &       & 1  & & 0.08	&   0.09   \\  
Zn\,{\sc i}   	& 0.03	& 0.05  & 2  & &  0.11 &  0.05  & 2  & &   0.11 & 0.05  &  2  & &  0.18 & 0.04   & 2  & &   0.18  &   0.09 & 2  & &  0.00 & 0.08  & 2  & &--0.05&   0.07   \\
Y\,{\sc ii}   	& 0.1	& 0.03  & 6  & &  0.16 &  0.02  & 6  & &   0.27 & 0.06  &  6  & &  0.35 &        & 1  & &   0.25  &        & 1  & &  0.30 &       & 1  & & 0.09	&   0.09   \\  
Zr\,{\sc i}   	& 001	&       & 1  & &  0.15 &        & 1  & &   0.23 &       &  1  & &  0.20 & 0.07   & 2  & &   0.05  &   0.07 & 2  & &  0.05 & 0.07  & 2  & &--0.03&   0.09   \\
Ba\,{\sc ii}    & 0.24	& 0.01  & 2  & &  0.31 &  0.02  & 2  & &   0.40 & 0.05  &  2  & &  0.30 & 0.05   & 2  & &   0.10  &   0.07 & 2  & &  0.10 & 0.05  & 2  & & 0.09	&   0.12   \\  
La\,{\sc ii}    & 0.11	&       & 1  & &  0.17 &        & 1  & &   0.29 &       &  1  & &  0.45 & 0.07   & 2  & &   0.29  &   0.01 & 2  & &  0.22 & 0.05  & 2  & & 0.11	&   0.12   \\  
Ce\,{\sc ii}    & 0.21	& 0.07  & 2  & &  0.30 &  0.06  & 2  & &   0.37 & 0.05  &  2  & &  0.30 &        & 1  & &   0.20  &        & 1  & &  0.15 &       & 1  & & 0.11	&   0.08   \\  
Nd\,{\sc ii}    & 0.33	& 0.05  & 3  & &  0.40 &  0.06  & 3  & &   0.35 & 0.03  &  3  & &  0.40 &        & 1  & &   0.40  &        & 1  & &  0.30 &       & 1  & & 0.22	&   0.04   \\  
Eu\,{\sc ii}    & 0.15	&       & 1  & &  0.15 &   	& 1  & &   0.20 &       &  1  & &  0.35 &        & 1  & &   0.25  &        & 1  & &  0.12 &       & 1  & & 0.06	&   0.09   \\  
                &      	&       &    & &       &        &    & &        &       &     & &       &        &    & &         &        &    & &  	  &       &    & & 	&          \\  
C/N             & 1.05	&       &    & &  1.48 &        &    & &   1.10 &       &     & &  1.41 &        &    & &   0.98  &        &    & &  1.38 &       &    && 1.23	 &  0.22   \\  
$^{12}$C/$^{13}$C& 6	&$\pm 1$&    & &  9    & $\pm 1$&    & &   12	&$\pm 1$&     & &  7	&$\pm 1$ &    & &    8	  & $\pm 1$&    & &   12  &$\pm 1$&    & & 9	&   2.5   \\  	
             \noalign{\smallskip}   
\hline
\hline
         \end{tabular}

\end{table}
\end{landscape}

\end{document}